\documentclass[runningheads]{llncs}

\usepackage[T1]{fontenc}
\usepackage{newtxtext}
\usepackage{newtxmath} 
\usepackage[scaled=.88,semibold,semicondensed]{noto-sans} 
\usepackage[scaled=.88,semibold,extracondensed]{noto-mono} 

\usepackage{xspace}
\usepackage[fleqn]{mathtools}
\usepackage{hyperref}

\usepackage{listings}

\let\LIN=\lstinline

\usepackage{tabularx}%
\newcolumntype{L}{>{$}l<{$}}%
\newcolumntype{C}{>{$}c<{$}}%
\newcolumntype{R}{>{$}r<{$}}%
\newcolumntype{M}{@{}p{\mathindent}@{}}%
\newcolumntype{t}{>{\mbox\bgroup}l<{\egroup}}%
\newcolumntype{e}{r@{\;}l}%
\newcolumntype{E}{>{$}r<{$}@{$\;$}>{$}l<{$}}%

\newcommand{\pbox}[2][c]{\begin{tabular}[#1]{@{}l@{}}#2\end{tabular}}

\newcommand{\Time}{\mathbb{T}}

\newcommand{\ALCIN}{\ensuremath{\cal ALCIN}\xspace}
\newcommand{\ALCIF}{\ensuremath{\cal ALCIF}\xspace}
\newcommand{\ALC}{\ensuremath{\cal ALC}\xspace}

\newcommand{\colonminus}{\mathrel{\text{:-}}}

\def\fun#1{\hbox{\sffamily\upshape #1}}

\def\kwfont{\upshape\sffamily\bfseries}
\newcommand{\kw}[1]{\text{\kwfont #1}}

\DeclareMathOperator{\NOT}{\kw{not}}
\DeclareMathOperator{\FAIL}{\kw{fail}}

\DeclareMathOperator{\LET}{\kw{let}}
\DeclareMathOperator{\CHOOSE}{\kw{choose}}
\DeclareMathOperator{\COLLECT}{\kw{collect}}
\DeclareMathOperator{\MATCH}{\kw{match}}
\DeclareMathOperator{\STH}{\kw{sth}}

\DeclareMathOperator{\var}{\mathit{var}}

\newcommand{\timevar}{\mathit{time}}
\DeclareMathOperator{\dom}{\mathit{dom}}

\RequirePackage{color}
\definecolor{DodgerUniformBlue}{rgb}{0.0,0.353,0.612}
\newcommand{\define}[1]{\emph{\textcolor{DodgerUniformBlue}{#1}}}

\begin{document}
\fontseries{\mddefault} 
\title{The Fusemate Logic Programming System \\ (System Description)}

\author{Peter Baumgartner\orcidID{0000-0002-6559-9654}}
\authorrunning{P. Baumgartner}

\institute{Data61/CSIRO and The Australian National University, Canberra, Australia
\href{mailto:Peter.Baumgartner@data61.csiro.au}{\email{Peter.Baumgartner@data61.csiro.au}}}
\maketitle              

\begin{abstract}
Fusemate is a logic programming system that implements the possible
model semantics for disjunctive logic programs. Its input language is centered around a
weak notion of stratification with comprehension and aggregation operators on top of it.
Fusemate is implemented as a shallow embedding in the Scala
programming language. This enables using Scala data
types natively as terms, a tight interface with external systems,
and it makes  model computation available as an ordinary container data structure
constructor.
The paper describes the above features and demonstrates them
with a non-trivial use-case,  the embedding of the description 
logic \ALCIF into Fusemate's input 
language.\\
\textbf{This version of the paper corrects an error in the published version, which used
  an unsuitable version of  ``blocking''  in the \ALCIF embedding.}
\end{abstract}

\section{Introduction}
\label{sec:introduction}
Fusemate\footnote{Fusemate is available at
  {\footnotesize\url{https://bitbucket.csiro.au/users/bau050/repos/fusemate/}}.} is a
logic programming system for computing possible models of disjunctive
logic
programs~\cite{Sakama:PossibleModelSemanticsDisjunctiveDatabases:DOOD:89,Sakama:Inoue:PossibleModels:JAR:94}.
A Fusemate logic program consists of (typically) non-ground
if-then rules with stratified default negation in the body~\cite{PRZYMUSINSKI1988193}.
Stratification entails that a 
true default-negated body literal remains true in the course of deriving new
conclusions.

Fusemate was introduced in~\cite{Baumgartner:PossibleModelsSpringer:IJCAR:2020} for
modelling systems that evolve over time and for analysing their current state based on the
events so far. Such tasks are often subsumed under the terms of stream processing,
complex event recognition, and situational awareness, and have been addressed (also) with
logic-based
approaches~\cite{DBLP:journals/ker/ArtikisSPP12,DBLP:journals/ai/BeckDE18,Baader:etal:SAIL:TABLEAUX:2009,DBLP:conf/ausai/BaaderBL15}.

To my knowledge, Fusemate is unique among all these and other logic programming
systems~\cite{DBLP:books/mc/18/Cat0BJD18,DBLP:conf/datalog/AlvianoFLPPT10,DBLP:journals/corr/GebserKKS14,DBLP:conf/lpnmr/SyrjanenN01,kozlenkov06edbt} (and theorem provers) in the way it is
implemented. Fusemate is implemented by shallow embedding in a full-fledged 
programming language, Scala~\cite{scala}.
Essentially, the user writes a syntactically sugared Scala program utilizing
familiar logic programming notation, and the program's execution returns models.
This has advantages and disadvantages. The main disadvantages is that it is more difficult
to implement performance boosting measures like term indexing.
The main advantage is that interfacing with data structure libraries and with external
systems is easy, an aspect whose importance has been emphasized for virtually all of the
above systems. In fact, 
Fusemate is motivated in parts by exploring how far the embedding approach can be pushed
and to what benefit.

The earlier Fusemate paper~\cite{Baumgartner:PossibleModelsSpringer:IJCAR:2020} focused on the
model computation calculus with a belief revision operator as the main novelty. It utilized
a certain notion of \emph{stratification by time (SBT)} for making the calculus effective
and useful in the intended application areas.
This system description focuses on the advantages of the shallow embedding approach as
boosted by new language features introduced here. These new language features are (a)
non-standard comprehension and aggregation operators, among others, and (b) a weaker
notion of \emph{stratification by time and predicates (SBTP)}.  In brief, SBTP is a
lexicographic combination of stratification by time and the standard stratification in
terms of the call-graph of the program. Section~\ref{sec:ALCIF} has an example that
demonstrates the need for (a) and (b) in combination, and Section~\ref{sec:usability}
discusses the shallow embedding approach and its advantages on a more general level.

Here is an excerpt from a Fusemate program that previews some of the new  features:
\begin{lstlisting}
type Time = java.time.LocalDateTime
val allIds = 1 to 10
case class Change(time:Time, id:Int, color:String) extends Atom 
case class State(time:Time, id:Int, color:String) extends Atom 
case class FullState(time:Time, drive:Set[Int], stop:Set[Int]) extends Atom
State(time, id, color) $\colonminus$ Now(time), CHOOSE(id:Int, allIds), Change(t <= time, id, color)
FullState(time, drive.toSet, stop.toSet) $\colonminus$ State(time,_,_),
  COLLECT(drive:List[Int], id STH State(time, id, "green")),
  COLLECT(stop:List[Int], id STH (State(time, id, color), color=="red" || color=="yellow"))
MovingState(time) $\colonminus$ FullState(time, drive, stop), stop.size < drive.size
Faulty(time, id, since) $\colonminus$  State(time, id, "red"),  Change(since < time, id, "green"),
  NOT (Change(t, id, "yellow"), since < t, t  < time)
\end{lstlisting}

The scenario comprises traffic lights identified by numbers 1 to 10 (line 2). In the course of time
the traffic lights change their colors, and each such event is recorded as a corresponding \fun{Change}
atom (line 3). The rule on line 6 computes a \fun{State} at a
current time \fun{Now(time)} as a snapshot of the current colors of all traffic lights. For that, the comprehension
\lstinline|Change(t <= time,id,color)| on line 6 finds the latest \fun{Change} event
before or at \fun{time} for a fixed \fun{id} chosen from \fun{allIds}, and binds that time
to the (unused) variable
\fun{t}.  A \fun{FullState}
aggregates the separate \fun{State} facts at a time partitioned as (Scala) sets of ids of
``drive'' and ``stop'' colors. In that, the \fun{\bfseries COLLECT} special form collects in a
Scala \fun{List}-typed variable the specified terms that satisfy the body behind \fun{\bfseries
  STH}. Notice that all atoms in \fun{FullState} refer to the same time, yet the program
is SBTP because \fun{State} comes before \fun{FullState}  in predicate stratification.
(Predicate stratification is computed automatically by Fusemate with Tarjan's algorithm.)
The rule on line 10 demonstrates the use of the Scala \fun{Set} method \fun{size}
in the body. Line 11 demonstrates the use of default negation in combination with comprehension.
When applied to a given sequence of \fun{Change} events, Fusemate computes 
models, one-at-a-time, each as  Scala set of atoms.

\section{Fusemate Programs}
\label{sec:programs}
For the purpose of this paper, a brief summary of the syntactic notions underlying Fusemate
programs is sufficient; see~\cite{Baumgartner:PossibleModelsSpringer:IJCAR:2020} for details.
Terms and atoms of a given signature are defined as usual. Let $\var(z)$ denote the set of
variables occurring in an expression $z$. We say that $z$ is \define{ground} if $\var(z) = \emptyset$.  
We write $z\sigma$ for applying a substitution $\sigma$ to
$z$. The domain of $\sigma$ is denoted by $\dom(\sigma)$.
A substitution $\gamma$ is a \define{grounding substitution for $z$} iff $\dom(\gamma) = \var(z)$
and $z\gamma$ is ground. In this case we simply say that \define{$\gamma$ is for $z$}.

Let $\Time$ be a countably infinite discrete set of \define{time points} equipped with a
total strict ordering $<$ (``earlier than''), e.g., the integers. Assume that the time
points, comparison operators $=$ and $\le$,  and a successor time function  $+ 1$ are part of
the signature and interpreted in the intended way.
A \define{time term} is a (possibly non-ground) term  over the sub-signature $\Time \cup
\{+1\}$.

The signature may contain other ``built-in'' predicate and function symbols for predefined types such as
strings, arithmetic data types, sets, etc.
We only informally assume that all terms are built in a well-sorted way and that built-in operators
over ground terms can be evaluated effectively. 

An \define{ordinary atom (with time term $t$)} is of the form $p(t, t_1,\ldots,t_n)$ where $p$ is an ordinary
predicate (i.e., neither a time predicate nor built-in), $t$ is a time term and $t_1,\ldots,t_n$ 
terms.  A \define{(Fusemate) rule} is an implication written in Prolog-like syntax as
    \begin{equation}
      \label{eq:rule}
      H \colonminus b_1,\ldots,b_k, \NOT \vec{b}_{k+1}, \ldots ,\NOT \vec{b}_n \enspace.
    \end{equation}
In~\eqref{eq:rule}, a rule \define{head} $H$ is either (a) a disjunction $h_1 \vee \cdots \vee h_m$ of ordinary atoms, for
    some $m \ge 1$, or (b) the expression $\FAIL$.\footnote{This definition of head is
      actually simplified as Fusemate offers an additional head 
      operator  for belief revision, see~\cite{Baumgartner:PossibleModelsSpringer:IJCAR:2020}. This is ignored here.} In case (a) the rule is
    \define{ordinary} and in case (b) it is a \define{fail rule}.
    A rule \define{body} $B$, the part to the
  right of $\colonminus$, is defined by mutual recursion as follows.
  A \define{positive body literal} is one of the following: (a) an ordinary atom, (b) a
  \define{comprehension atom (with time term $x$)} of the form
  $p(x \circ t, t_1,\ldots,t_n) \STH B$, where $x$ is a variable,
  $\circ \in \{<, \le, >, \ge \}$ and $B$ is a body, (c) a built-in call , i.e., an atom with a
  built-in predicate symbol, or (d) a \define{special form} $\LET(x, t)$,
  $\CHOOSE(x, \mathit{ts})$, $\MATCH(t, s)$ or $\COLLECT(x, t \STH B)$ where $x$ is a
  variable, $s,t$ are terms, $\mathit{ts}$ is a list of terms, and $B$ is a body.
  A \define{positive body} is a list $\vec{b} = b_1,\ldots,b_k$ of positive body literals with
  $k \ge 0$.  If $k = 0$ then $\vec{b}$ is \define{empty} otherwise it is
  \define{non-empty}. 
  A \define{negative body
    literal} is an expression of the form $\NOT \vec{b}$, where $\vec{b}$ is a non-empty
  positive body.  A \define{body} is a list
  $B = b_1,\ldots,b_k, \NOT \vec{b}_{k+1}, \ldots ,\NOT \vec{b}_n$ comprised of a (possibly empty)
  positive body and (possibly zero) negative body literals. It is \define{variable free}
  if $\var(b_1,\ldots,b_k) = \emptyset$.

Let $r$ be a rule~\eqref{eq:rule}.
We say that $r$ is \define{range-restricted} iff $\var(H) \subseteq \var(\vec{b})$. Compared to
the usual notion of range-restrictedness~\cite{Manthey:Bry:SATCHMO:88}, Fusemate rules may contain extra variables
in negative body literals. For example, $\fun{p}(t, x) \colonminus \fun{q}(t, x), \NOT (s <
t,\fun{r}(s, x, y) )$ is range-restricted in our sense with extra variables $s$ and
$y$. The extra variables are implicitly existentially quantified within the $\NOT$
expression. The example corresponds to the formula $ \fun{q}(t, x) \land \neg \exists s, y . (s < t \land
\fun{r}(s, x, y)) \rightarrow \fun{p}(t, x)$. Semantically and operationally this will cause no
problems thanks to stratification, introduced next.

Fusemate programs -- sets of rules -- need to be ``stratified by time and by predicates'' (SBTP). The standard
notion of stratification by predicates means that the call graph of the 
program contains no cycles going through negative body literals. The edges of this
 call graph are the ``depends on'' relation between predicate symbols such that $p$
 positively (negatively) depends on $q$ if there is a rule with a $p$-atom in its head
 and a $q$-atom in its positive (negative) body. For disjunctive heads, all head
 predicates are defined to depend positively on each other.  Every strongly connected
 component of the call graph is called a stratum, and in predicate stratified programs negative body literals can
 occur only in strata lower than the head stratum.

SBTP is defined as follows:
 for every rule \eqref{eq:rule} in a given program, (a) there is a variable $\timevar$ that is
    the time term of some ordinary $b \in \vec{b}$,
(b)     if $H$  is an ordinary head then every head literal must have a time term
constrained to be $\ge$ than $\timevar$, and (c) for all rule bodies $B$ occurring in the rule:
    \begin{enumerate}
      \item[(i)] the time term of every ordinary or comprehension body literal
        in $B$ must be constrained to be $\le$ than $\timevar$, and
      \item[(ii)] for every negative body literal $\NOT \vec{b}$  in $B$ (including the
        top-level body of \eqref{eq:rule}) and every
          ordinary or comprehension literal $b \in \vec{b}$ , 
        the time term of $b$ must constrained to be (i) $<$ than $\timevar$ or (ii)
        $\le$  than $\timevar$ and
        the predicate symbol of $b$ is in a lower stratum than $H$. 
      \end{enumerate}
      For the purpose of this paper we only informally assume that all rules contain constraints for enforcing the required time ordering
      properties. There are similar stratification requirements for comprehension atoms
      and special forms so that their evaluation satisfies the counterpart of condition (ii) (see
      below for $\COLLECT$). A fully formal definition could be given by modifying the 
    spelled-out definition of
    SBT in~\cite{Baumgartner:PossibleModelsSpringer:IJCAR:2020}.

    As an example, if $\fun{r}$ belongs to a lower stratum than $\fun{p}$ then the following
    five rules all are SBTP, while only the first two rules are SBT.
    \begin{align}
      \fun{p}(\timevar, x) & \colonminus \fun{q}(\timevar, x), \fun{r}(t, y), t\le\timevar\\
      \fun{p}(\timevar, x) & \colonminus \fun{q}(\timevar, x), \NOT ( \fun{r}(t, y), t<\timevar )\\
      \fun{p}(\timevar, x) & \colonminus \fun{q}(\timevar, x), \NOT ( \fun{r}(t, y), t\le\timevar )\\ 
      \fun{p}(\timevar +1, x) & \colonminus \fun{q}(\timevar, x), \NOT ( \fun{r}(t, y), t\le\timevar)\\
      \fun{p}(\timevar, x) & \colonminus \fun{q}(\timevar, x), (\fun{p}(t < \timevar, y) \STH
                          q(t, y) ), \fun{r}(t, y)
    \end{align}
    Finally, a \define{(Fusemate) program} is  a set of range-restricted rules that is SBTP.

\section{Model Computation}
    \label{sec:model-computation}
The possible model semantics of disjunctive logic
programs~\cite{Sakama:PossibleModelSemanticsDisjunctiveDatabases:DOOD:89,Sakama:Inoue:PossibleModels:JAR:94}
associates to a given disjunctive program a certain set of normal programs (i.e., without
disjunctive heads) and takes the intended model(s) of these normal programs as the possible models
of the given program. These ``split'' programs represent all possible ways of making one or
more head literals true, for every disjunctive rule.  As a propositional example, the 
program $\{ a \colonminus b,\ a \vee c \colonminus b, b \colonminus \ \} $ is associated to the split programs
$\{ a \colonminus b, b \colonminus \ \} $ and
$\{ a \colonminus b,\ c \colonminus b, b \colonminus \ \} $. The possible models, hence, are
$\{a, b \}$ and $\{a, b, c \}$

Fusemate computes possible models by bottom-up fixpoint computation and dynamic grounding
the program rules in the style of hyper
tableaux~\cite{Baumgartner:Furbach:Niemelae:HyperTableau:JELIA:96}. The model computation
procedure is implemented as a variant of the well-known given-clause algorithm, which
seeks to avoid deriving the same conclusion from the same premises twice. It exhausts
inferences in an outer loop/inner loop fashion according to the given program's
stratification by time and by predicates. The main data structure is a set of paths,
where each path represents a partial model candidate computed so far
(see~\cite{Baumgartner:PossibleModelsSpringer:IJCAR:2020} for more details).
Paths are selected, extended,
split and put back into the set until exhausted, for a depth-first, left-to right
inference strategy. Paths carry full status information, which is instrumental for 
implementing incrementality, such that facts with current or later time can be added at any stage
without requiring model recomputation from scratch. This, however, necessitated keeping
already exhausted paths for continued inferences later. 

The proof procedure's core operation is computing a \emph{body matcher}, i.e., a substitution $\gamma$ for a rule's
positive body variables so that the rule body becomes satisfied in the current
partial model candidate. 
Formally, let $I$ be a set of ordinary ground atoms, representing the obvious interpretation
that assigns true to exactly the members of $I$. Let $B$ be a body.
A \define{body matcher for $B$} is a substitution $\gamma$ for the positive body of $B$ , 
written as  $I, \gamma \models B$, such that the following holds ($b, B$ means the sequence of head $b$ and rest body $B$):

{\renewcommand{\arraystretch}{1.2}
\begin{tabular}{RLRL}
  I, \varepsilon & \models \epsilon && \text{($\epsilon$ is the empty body and $\varepsilon$ is the empty substitution)} \\
  I,\gamma\sigma & \models b, B & \quad & \text{iff }\pbox[t]{$\gamma$ is for $b$, $b\gamma \in I$ and $I,\sigma  \models B\gamma$,
     with $b$  ordinary atom}\\
  I,\gamma\sigma & \models \mathrlap{(p(x \stackrel{\le}{<} t, t_1,\ldots,t_n) \STH C), B \text{ iff $\gamma$ is for $p(x, t_1,\ldots,t_n)$ and}} \\
       &  &  & {\renewcommand{\arraystretch}{1.0}\pbox[t]{
               (1) $p(x, t_1,\ldots,t_n)\gamma \in I$,  $x\gamma < t$ and  $I,\delta \models C\gamma$ for some $\delta$,\\
    (2) there is no $\gamma'$ for $p(x, t_1,\ldots,t_n)$ and no $\delta$ such that \\ \qquad $p(x, t_1,\ldots,t_n)\gamma' \in I$, 
 $x\gamma < x\gamma' \stackrel{\le}{<} t$  and $I,\delta \models C\gamma'$, and \\
    (3)  $I,\sigma  \models B\gamma$  }} \\
I,\sigma & \models a, B & & \text{iff $a$ evaluates to true and $I,\sigma \models B$ where $a$ is ground built-in}\\
I,\gamma\sigma & \models \LET(x, t), B & & \text{iff $\gamma = [x \mapsto t]$ and $I,\sigma \models B\gamma$} \\
I,\gamma\sigma & \models \mathrlap{\CHOOSE(x, \mathit{ts}), B \text{ iff $\gamma = [x \mapsto t]$ and $I,\sigma \models B\gamma$ for some $t \in \mathit{ts}$}} \\
  I,\gamma\sigma & \models \mathrlap{\MATCH(t, s), B \text{ iff $\gamma$ is for $t$,  $t\gamma = s$ and $I,\sigma \models B\gamma$}} \\
  I,\gamma\sigma & \models \mathrlap{\COLLECT(x, t \STH C), B \text{ iff $\gamma = [x \mapsto \{t\delta \mid I,\delta \models C \}]$ and $I,\sigma \models B\gamma$}} \\
  I,\sigma & \models \NOT \vec{b}, B & \quad & \text{iff there is no $\delta$ such    that $I,\delta \models \vec{b}$, and $A,\sigma  \models B$}
\end{tabular}
}

A \define{comprehension atom} $p(x \circ t, t_1,\ldots,t_n) \STH B$  stands for the subset of all
ground $p$-instances in $I$ such that $B$ is satisfied and with a time $x$ as
close as possible to $t$ wrt.\ $<$ or $\le$. The cases  for $>$ and $\ge$ are dual and not spelled out above to save 
space. The $\COLLECT$ special form collects in the variable $x$ the set of all instances
of  term $t$ such that the body $C$ is satisfied in $I$. 
We require comprehension atoms and $\COLLECT$s to be used in a
stratified way, so that their results do not change later in a derivation when $I$ is
extended. The requirements are the same as with $\NOT$ and can
be enforced by ordering constraints.

The definition above extends the earlier definition of body matchers
in~\cite{Baumgartner:PossibleModelsSpringer:IJCAR:2020} with the new comprehension construct
and the $\LET$, $\CHOOSE$, $\MATCH$, $\COLLECT$ operators.  It now also enforces 
left-to-right evaluation of $B$ because the new binding operators
depend on a fixed order guarantee to be useful.
An example is the (nonsensical) body
\LIN{CHOOSE (x: Int, List(1, 2, 3)), LET(xxx: Int,  3*x), xxx 
which relies on this order. Undefined cases, e.g., when
evaluation of a non-ground built-in is attempted, or when a 
binder variable has already been used before are detected
as compile time syntax errors.

\section{Shallow Embedding in Scala}
\label{sec:usability}
Fusemate is implemented as a shallow embedding into Scala~\cite{scala}. 
It has three conceptual main components: a signature framework,
a Scala compiler plugin, and an inference engine for fixpoint computation as explained in
Section~\ref{sec:model-computation}.
The signature framework provides a set
of Scala class definitions as the syntactical basis for writing Fusemate programs. It is
parameterized in a type \fun{Time}, which can be any Scala or Java type that is equipped
with an ordering and an addition function for time increments, for example \LIN{Int} or
\LIN{java.time.OffsetDataTime}. The programmer then refines an abstract class
\fun{Atom} of the \fun{Time}-instantiated signature framework with definitions of
predicate symbols and their (Scala-)sorted arities. See lines (3)-(5) in the 
program in the introduction for an example. These atoms then can be used in Fusemate rules,
see lines (6)--(12) in the example.

While written in convenient syntax, rules are syntactically ill-formed Scala.
This problem is solved by the compiler plugin, which intercepts the compilation
of the input file at an early stage and transforms the rules into valid Scala source code.\footnote{Early
  experiments showed it is cumbersome and
  error-prone to write the Scala code by hand, so this was not an option.
  The compiler plugin is written in Scala and operates at the abstract syntax tree level. This was
  conveniently be done thanks to a sophisticated quasiquote mechanism.}
More precisely, a rule is transformed into
a curried partial function that is parameterized in an interpretation context \LIN{I}.
The curried parameters are Scala guarded pattern matching expression
and correspond to the rule's positive
body literals, in order.  For example, the \LIN{Faulty} rule on lines
(11) and (12), with the condition \LIN{since < time} ignored, for simplicity,  is (roughly)
translated into the function $f$
\begin{lstlisting}
(I: Interpretation) => { case State(time, id, "red") => {
 case Change(since, id1, "green") if id == id1 &&
  ({ case Change(t, id2, "yellow") if id == id2 && since < t && t < time => FAIL} failsOn I) =>
     Faulty(time, id, since) } }
\end{lstlisting}
Notice the renaming of repeated occurrences of the \LIN{id} variable, which is needed for
the correct semantics. Notice also that a Scala Boolean-valued expression in an ordinary
body literal position (e.g., \LIN{t <  time}) simply becomes a guard in a pattern. 

The code above can be understood with body matcher computation in mind. Suppose the inference
engine selects an interpretation $I$ from the current set of paths. For exhausting
$f$ on $I$, the inference engine combinatorially chooses literals $l_1, l_2 \in I$ and
collects the evaluation results of $f(I)(l_1)(l_2)$, if defined.
Observe that by the transformation into Scala pattern matching,
body matchers are only implicitly computed by the Scala
runtime system. Each evaluation result, hence, is a
body-matcher instantiated head.

The rule's negative body literal is translated into the code on line (3) and conjoined
to the guard of the preceding ordinary literal. In general, a
negative literal \LIN{NOT} $\mathit{body}$ is treated by translating \LIN{FAIL} $\colonminus$ \LIN{NOT}
$\mathit{body}$ and evaluating the resulting Scala code on $I$ by means of the
\LIN{failsOn} method. If 
\LIN{FAIL} is not derivable then \LIN{NOT} $\mathit{body}$ is satisfied. Again,
appropriate bindings for the variables bound outside of $\mathit{body}$ are held
implicitly by the Scala runtime system. 
The translation of the special forms and comprehension is not explained here for space
reasons. Fusemate can show the generated code, though.

\subsection*{Properties and Advantages}
The shallow embedding approach enables introspection capabilities and interfacing between
the rule language and the host language beyond what is implemented in other systems. In
Fusemate, the terms of the logical language are nothing but Scala objects.  As a consequence, any
available Scala type or library data structure can be used as a built-in without extending
an ``interface'' to an extension language -- simply because there is none. Dually, the
embedding of the rule language into the host language Scala is equally trivial because rules,
atoms and interpretations are Scala objects, too. 

It is this ``closed loop'' that makes an aggregation operator ($\COLLECT$)  possible that returns a
list of Scala objects as specified by the programmer, e.g., a list of terms or
atoms.\footnote{Technically, this is possible because the current interpretation is
  available in the rule body through the parameter \LIN{I} (see the transformation example
  above).  One could directly access \LIN{I}, e.g., as in
  \LIN{CHOOSE(a: atom, I), MATCH(State(t,3,c), a), t>10, c != "red"}} This list can be further analysed or
manipulated by the rules. See the description logic embedding in
Section~\ref{sec:ALCIF}, which critically depends on this feature.
This introspection capability stands out in comparison to the logic programming
systems mentioned in the introduction. For instance, aggregation in systems like
DLV~\cite{DBLP:conf/datalog/AlvianoFLPPT10},  and IDB~\cite{DBLP:books/mc/18/Cat0BJD18}
is limited to predefined integer-valued aggregates for sum, count, times, max and min.

Most logic programming systems can be called from a (traditional) host programming
language and can call external systems or utilize libraries for data structures. The DLV
system, for instance, interfaces with C++ and Python~\cite{dlvhex-journal:2016}, Prova~\cite{kozlenkov06edbt} with Java, and
IDP with the Lua scripting language. Systems based on grounding (e.g., DLV and IDP) face
the problem of ``value invention'' by external calls, i.e., having to deal with terms that
are not part of the input specification~\cite{DLV-value-invention:2007}. 

The main issue, however, from the Fusemate perspective is that these systems' external
interfaces are rather heavy-handed (boilerplate code, mapping logic terms 
to/from the host language, String representation of logic programs) and/or limited to a
predefined set of data structures. 
In contrast, Fusemate's seamless  integration with Scala encourages
a more integrated and experimental problem solving workflow.
The following Scala program demonstrates this point with the traffic light example:
\begin{lstlisting}[mathescape=false]
List("2020-07-02T10:00:00,1,green", .., "2020-07-02T10:02:15,2,red")
      .map { _.split(",") } // Split CSVs intos triple, represented as Java array
      .map { // Convert String triple to positive Change literals
        case Array(date,id,color) => Change(LocalDateTime.parse(date), id.toInt, color) }
      .saturate { rules } // saturate is the Fusemate call, computes all models of the rules
      .head  // Select the first model
      .toList // Convert to Scala List because we want to sort elements by time:
      .sortBy { _.time }
      .flatMap { // Analyze literals in model and retain only Faulty ones as CSV
        case Faulty(time, id, since) => List(s"$time,$id,$since")
        case _ => List() }
\end{lstlisting}
From a workflow perspective, this program integrates Fusemate as a list operator (on a list of
\LIN{Change} instances) in an otherwise unremarkable functional program. 

For a more realistically sized experiment I tried a combined Fusemate/Scala
workflow for analysing the data of the DEBS 2015 Grand
Challenge.\footnote{\url{http://www.debs2015.org/call-grand-challenge.html}}
The data comprises two millions taxi rides in New York City in terms of start/end times, and
start/end GPS coordinates, among others. The problem considered was to detect anomalies where a taxi driver 
drivers away from a busy hotspot without a passenger. 
Solving the problem required clustering locations by pickup/drop-off
activity for determining hotspots, and then analysing driver behavior given their
pickups/drop-offs at these hotspots.

Two million data points were too much for Fusemate alone and required Scala preprocessing,
e.g., for filling a grid abstraction of New York coordinates, data cleansing and filtering out
little active drivers. Fusemate was used for
computing clusters with rules similar to transitive closure computation.  Input to
Fusemate calls were Scala precomputed point clouds.
The computed clusters were used to analyze Scala prefiltered taxi rides for anomaly
detection based on the clusters. This involved three moderately complex rules,
for first identifying gaps and then analysing them.
The comprehension operator was useful to find ``the most recent ride predating a given
start'', among others. The longest Fusemate run was 0.31sec for 64 rides (with 39
clusters fixed), most other runs took less than 0.15sec. Fusemate's performance was
perfectly acceptable in this experiment thanks to a \emph{combined} workflow. 
\section{Embedding Description Logic \ALCIF}
\label{sec:ALCIF}
\ALCIF is the well-known description logic \ALC extended with inverse roles
and functional roles. (See~\cite{Baader:etal:DLHandbookBook:2002} for background on
description logics.)
This section describes how to translate an \ALCIF knowledge base to Fusemate rules and
facts for satisfiability checking.

This is our example knowledge base, TBox on the left, ABox on the right:
{
\setlength{\jot}{1pt}
  \begin{xalignat*}{2}
  \mathsf{Person} & \sqsubseteq \mathsf{Rich} \sqcup \mathsf{Poor}  & \qquad  \mathsf{Anne} &:\mathsf{Person} \sqcap \mathsf{Poor}\\
  \mathsf{Person} & \sqsubseteq \exists \mathsf{father} .  \mathsf{Person}  &  (\mathsf{Anne},   \mathsf{Fred}) &: \mathsf{father} \\
  \mathsf{Rich} & \sqsubseteq \forall \mathsf{father}^{-1} .  \mathsf{Rich} & \mathsf{Bob}  &:\mathsf{Person} \\
  \mathsf{Rich} \sqcap \mathsf{Poor} & \sqsubseteq \bot &  (\mathsf{Bob},   \mathsf{Fred}) &: \mathsf{father} 
\end{xalignat*}
}
The $\mathsf{father}$ role is declared as functional, i.e., as a right-unique relation, and
$\mathsf{father}^{-1}$ denotes its inverse ``child'' relation.
The third GCI says that all children of a rich father are rich as
well.
In all models of the knowledge base $\mathsf{Fred}$ is $\mathsf{Poor}$. This
follows from the given fact that his child $\mathsf{Anne}$ is poor, functionality of
$\mathsf{father}$ and the third CGI.
However, there are models where $\mathsf{Bob}$  is $\mathsf{Rich}$ and models where $\mathsf{Bob}$  is $\mathsf{Poor}$.

  Translating description logic into rule-based languages has been done in many ways, see e.g.~\cite{DBLP:journals/jair/MotikSH09,DBLP:journals/tplp/LukacsyS09,DBLP:conf/www/GrosofHVD03,DBLP:conf/ijcai/CarralK20}.
  An obvious starting point is taking the FOL version of a given knowledge base.
  Concept names become unary predicates, role names become binary
predicates, and GCIs (general concept inclusions) are translated into implications. By polynomial transformations, the
implications can be turned into clausal form (if-then rules over literals), except for
existential quantification in a positive context, which causes unbounded Skolem
terms in derivations when treated naively (for example, the third CGI above is problematic
in this sense). This is why many systems and also the transformation to Fusemate below
avoid Skolemization.

The first GCI corresponds to the clause
$\mathsf{Person}(x) \to \mathsf{Rich}(x) \vee \mathsf{Poor}(x)$, and the second
corresponds to the ``almost'' clause $\mathsf{Person}(x) \to \exists y . (\mathsf{father}(x,y) \land \mathsf{Person}(y))$.
Fusemate works with the reified rule versions of these, with an $\mathsf{IsA}$-predicate for concept instances, and a
$\mathsf{HasA}$-predicate for role instances.  For the whole TBox one
obtains the following, where \fun{RN} stands for ``role name'' and \fun{CN} stands for
``concept name''.\footnote{See the Fusemate web page for the full, runnable code.}
{\small
\begin{lstlisting}
IsA(x, Exists(RN("father"), CN("Person")), time) $\colonminus$  IsA(x, CN("Person"), time)
IsA(x, CN("Rich"), time) OR IsA(x, CN("Poor"), time) $\colonminus$ IsA(x, CN("Person"), time)
IsA(x, Forall(Inv(RN("father")), CN("Rich")), time) $\colonminus$ IsA(x, CN("Rich"), time)
FAIL $\colonminus$ IsA(x, CN("Poor"), time), IsA(x, CN("Rich"), time)
functionalRoles = Set(RN("father"))
\end{lstlisting}
  }
Every GCI can be converted into rules like the above without problems. For that, starting from its
NNF, $\exists$-quantifications in the premise of a rule can be expanded in place, and
$\forall$-quantifications can be moved to the head as the $\exists$-quantification of the NNF
of the negated formula. Similarly for negated concept names. See~\cite{DBLP:journals/jair/MotikSH09} for such transformation methods.
The ABox is represented similarly. Its first element, for instance, is \LIN{IsA(Name("Anne"), And2( CN("Person"), CN("Poor")), 0)}.

In addition, some more general ``library'' rules for the tableau calculus are needed:
{\small
\begin{lstlisting} 
IsA(x, c1, time) AND IsA(x, c2, time) $\colonminus$ IsA(x, And2(c1, c2), time)
IsA(x, c1, time) OR IsA(x, c2, time) $\colonminus$ IsA(x, Or2(c1, c2), time)
// Expansion rules for quantifiers
IsA(y, c, time) $\colonminus$ Neighbour(x, r, y, time), IsA(x, Forall(r, c), time)
HasA(x, r, rSuccOfx, time+1) AND IsA(rSuccOfx, c, time+1): @preds("TimePlus1") $\colonminus$
  IsA(x, Exists(r, c), time),  ! (functionalRoles contains r),
  NOT( Neighbour(x, r, y, time), IsA(y, c, time) ),  NOT( Blocked(x, _, time) ),
  LET(rSuccOfx: Individual, Succ(r, x))

HasA(x, r, rSuccOfx, time+1) AND IsA(rSuccOfx, c, time+1): @preds("TimePlus1") $\colonminus$ (
  IsA(x, Exists(r, c), time),  functionalRoles contains r,
  NOT( Neighbour(x, r, y, time) ),  NOT( Blocked(x, _, time) ),
  LET(rSuccOfx: Individual, Succ(r, x))

IsA(y, c, time) $\colonminus$
  IsA(x, Exists(r, c), time),  functionalRoles contains r,
  Neighbour(x, r, y, time)
\end{lstlisting}
}
The expansion rules on lines 1 and 2 deal with the $\ALC$ binary Boolean connectives $\fun{And2}$ and
$\fun{Or2}$ in the obvious way. Supposing NNF of embedded formulas, no other cases can apply.
The remaining rules can be understood best with the standard
tableau algorithm for \ALCIN in mind, which includes blocking to guarantee termination. They follow the terminology in~\cite[Chapter 4]{DBLP:books/daglib/0041477}.
The $\fun{Neighbour}$ relation abstracts from the $\fun{HasA}$ relation, left away for
space reasons. The expansion rule for $\exists$ comes for three cases. The first case
(line 5), for example,  applies to non-functional roles as per the Scala builtin test on line 6. The
expansion of the given
$\exists$-formula only happens if it is not yet satisfied and in a non-blocked
situation (line 7). In this case the rule derives a Skolem object defined on line 8 for
satisfying the $\exists$-formula. Notice the annotation \LIN{@preds("TimePlus1")}
which makes sure that the head is on the highest stratum. This way, the rule will be
applied after, in particular,  the rules for blocking. 
Furthermore, with the time stamp \fun{time +1} the Skolem object is kept separate from the
computations in the current iteration \fun{time}.
The  blocking rules are defined as follows:
{\small
\begin{lstlisting} 
// Collect all concepts that an individual x isA, at a given time
Label(x, cs.toSet, time) $\colonminus$ IsA(x, _, time), COLLECT(cs: List[Concept], c STH IsA(x, c, time))
// Ancestor relation of Skolem objects introduced by exists-right
Anc(x, Succ(r, x), time) $\colonminus$ HasA(x, r, Succ(r, x), time)
Anc(x, Succ(r, z), time) $\colonminus$ HasA(z, r, Succ(r, z), time), Step(time, prev), Anc(x, z, prev)
// Blocked case 1: y is blocked by some individual x according to 'pairwise ancestor blocking'
Blocked(y, x, time) $\colonminus$ Anc(x, y, time), Label(y, yIsAs, time), Label(x, xIsAs,time), yIsAs == xIsAs,
  HasA(y1, r, y, time), HasA(x1, r, x, time), Label(y1, y1IsAs, time), Label(x1, x1IsAs, time), y1IsAs == x1IsAs
// Blocked case 2: y is blocked by some ancestor
Blocked(y, x, time) $\colonminus$ Anc(x, y, time), Blocked(x, _, time)
\end{lstlisting}
  }
Some additional rules are needed for dealing with basic inconsistencies and for carrying over
$\fun{IsA}$ and  $\fun{HasA}$ facts between iterations. They are not shown here.

The expansion rules and blocking rules follow the tableau calculus description in~\cite[Chapter
4]{DBLP:books/daglib/0041477}. 
One important detail is that the expansion rule
for $\exists$ must be applied with lowest priority. This is straightforward thanks
to Fusemate's stratification and aggregation construct. Equally important
is the access to (Scala) data structures via built-ins and using them as terms
of the logical language. This made it easy to program Skolemization and the $\fun{Label}$
relation for collecting sets of concepts of an individual.

\section{Conclusions}
This paper described recent developments around the Fusemate logic programming system.  It
included new technical improvements for a weaker form of stratification, which enabled
useful aggregation and comprehension language constructs. It also argued for the
advantages of the tight integration with Fusemate's host language, Scala, in terms of data
structures and usability.

Answer set solvers like DLV and SModels are designed to solve NP-complete or higher
complexity search problems as fast as possible. Fusemate is not motivated as a
competitive such system, it is motivated for "well-behaved" knowledge representation
applications, similarly to description logic reasoners, whose (often) NExpTime complete
solving capabilities are not expected to be typically needed.
(Some more work is needed, though, e.g., on improving the current term indexing techniques
to speed up model computation.)
More specifically, the main intended application of Fusemate is for the runtime analysis of systems that
evolve over time. The taxi rides data experiment explained in Section~\ref{sec:usability}
is an example for that. It suggests that Fusemate is currently best used in a combined
problem solving workflow if scalability is an issue.

As for future work, the next steps are to make the description logic reasoner of
Section~\ref{sec:ALCIF} callable from within Fusemate rules in a
DL-safe way~\cite{10.1007/978-3-540-30475-3_38} and to embed a temporal reasoning formalism. 
The event calculus~\cite{DBLP:journals/ngc/KowalskiS86} seems to be a good fit.

\paragraph{Acknowledgements.} I am grateful to the reviewers for their helpful comments.
\enlargethispage{2ex}

  \clearpage\newpage
  
\bibliographystyle{splncs04}

\begin{thebibliography}{10}
\providecommand{\url}[1]{\texttt{#1}}
\providecommand{\urlprefix}{URL }
\providecommand{\doi}[1]{https://doi.org/#1}

\bibitem{DBLP:conf/datalog/AlvianoFLPPT10}
Alviano, M., Faber, W., Leone, N., Perri, S., Pfeifer, G., Terracina, G.: The
  disjunctive datalog system {DLV}. In: de~Moor, O., Gottlob, G., Furche, T.,
  Sellers, A.J. (eds.) Datalog Reloaded - First International Workshop, Datalog
  2010, Oxford, UK, March 16-19, 2010. Revised Selected Papers. Lecture Notes
  in Computer Science, vol.~6702, pp. 282--301. Springer (2010).
  \doi{10.1007/978-3-642-24206-9\_17},
  \url{https://doi.org/10.1007/978-3-642-24206-9\_17}

\bibitem{DBLP:journals/ker/ArtikisSPP12}
Artikis, A., Skarlatidis, A., Portet, F., Paliouras, G.: Logic-based event
  recognition. Knowl. Eng. Rev.  \textbf{27}(4),  469--506 (2012).
  \doi{10.1017/S0269888912000264},
  \url{https://doi.org/10.1017/S0269888912000264}

\bibitem{Baader:etal:DLHandbookBook:2002}
Baader, F., Calvanese, D., McGuinness, D., Nardi, D., Patel-Schneider, P.
  (eds.): Description Logic Handbook. Cambridge University Press (2002)

\bibitem{Baader:etal:SAIL:TABLEAUX:2009}
Baader, F., Bauer, A., Baumgartner, P., Cregan, A., Gabaldon, A., Ji, K., Lee,
  K., Rajaratnam, D., Schwitter, R.: A novel architecture for situation
  awareness systems. In: Giese, M., Waaler, A. (eds.) Automated Reasoning with
  Analytic Tableaux and Related Methods (TABLEAUX 2009). LNAI, vol.~5607, pp.
  77--92. Springer (July 2009). \doi{10.1007/978-3-642-02716-1\_7},
  \url{SAIL-TABLEAUX-09.pdf}

\bibitem{DBLP:conf/ausai/BaaderBL15}
Baader, F., Borgwardt, S., Lippmann, M.: Temporal conjunctive queries in
  expressive description logics with transitive roles. In: Pfahringer, B.,
  Renz, J. (eds.) {AI} 2015: Advances in Artificial Intelligence - 28th
  Australasian Joint Conference, Canberra, ACT, Australia, November 30 -
  December 4, 2015, Proceedings. Lecture Notes in Computer Science, vol.~9457,
  pp. 21--33. Springer (2015). \doi{10.1007/978-3-319-26350-2\_3},
  \url{https://doi.org/10.1007/978-3-319-26350-2\_3}

\bibitem{DBLP:books/daglib/0041477}
Baader, F., Horrocks, I., Lutz, C., Sattler, U.: An Introduction to Description
  Logic. Cambridge University Press (2017),
  \url{http://www.cambridge.org/de/academic/subjects/computer-science/knowledge-management-databases-and-data-mining/introduction-description-logic?format=PB\#17zVGeWD2TZUeu6s.97}

\bibitem{Baumgartner:PossibleModelsSpringer:IJCAR:2020}
Baumgartner, P.: {Possible Models Computation and Revision -- A Practical
  Approach}. In: Peltier, N., Sofronie-Stokkermans, V. (eds.) International
  Joint Conference on Automated Reasoning. LNAI, vol. 12166, pp. 337--355.
  Springer International Publishing, Cham (2020).
  \doi{10.1007/978-3-030-51074-9\_19}, \url{possible-models-IJCAR-2020.pdf}

\bibitem{Baumgartner:Furbach:Niemelae:HyperTableau:JELIA:96}
Baumgartner, P., Furbach, U., Niemel{\"a}, I.: {Hyper Tableaux}. In: Logics in
  Artificial Intelligence (JELIA '96). No.~1126 in Lecture Notes in Artificial
  Intelligence, Springer (1996), \url{tableaux-jelia-llncs.pdf}

\bibitem{DBLP:journals/ai/BeckDE18}
Beck, H., Dao{-}Tran, M., Eiter, T.: {LARS:} {A} logic-based framework for
  analytic reasoning over streams. Artif. Intell.  \textbf{261},  16--70
  (2018). \doi{10.1016/j.artint.2018.04.003},
  \url{https://doi.org/10.1016/j.artint.2018.04.003}

\bibitem{DLV-value-invention:2007}
Calimeri, F., Cozza, S., Ianni, G.: External sources of knowledge and value
  invention in logic programming. Annals of Mathematics and Artificial
  Intelligence  \textbf{50},  333--361 (08 2007).
  \doi{10.1007/s10472-007-9076-z}

\bibitem{DBLP:conf/ijcai/CarralK20}
Carral, D., Kr{\"{o}}tzsch, M.: Rewriting the description logic {ALCHIQ} to
  disjunctive existential rules. In: Bessiere, C. (ed.) Proceedings of the
  Twenty-Ninth International Joint Conference on Artificial Intelligence,
  {IJCAI} 2020. pp. 1777--1783. ijcai.org (2020).
  \doi{10.24963/ijcai.2020/246}, \url{https://doi.org/10.24963/ijcai.2020/246}

\bibitem{DBLP:books/mc/18/Cat0BJD18}
Cat, B.D., Bogaerts, B., Bruynooghe, M., Janssens, G., Denecker, M.: Predicate
  logic as a modeling language: the {IDP} system. In: Kifer, M., Liu, Y.A.
  (eds.) Declarative Logic Programming: Theory, Systems, and Applications, pp.
  279--323. {ACM} / Morgan {\&} Claypool (2018). \doi{10.1145/3191315.3191321},
  \url{https://doi.org/10.1145/3191315.3191321}

\bibitem{DBLP:journals/corr/GebserKKS14}
Gebser, M., Kaminski, R., Kaufmann, B., Schaub, T.: Clingo = {ASP} + control:
  Preliminary report. CoRR  \textbf{abs/1405.3694} (2014),
  \url{http://arxiv.org/abs/1405.3694}

\bibitem{DBLP:conf/www/GrosofHVD03}
Grosof, B.N., Horrocks, I., Volz, R., Decker, S.: Description logic programs:
  combining logic programs with description logic. In: Hencsey, G., White, B.,
  Chen, Y.R., Kov{\'{a}}cs, L., Lawrence, S. (eds.) Proceedings of the Twelfth
  International World Wide Web Conference, {WWW} 2003, Budapest, Hungary, May
  20-24, 2003. pp. 48--57. {ACM} (2003). \doi{10.1145/775152.775160},
  \url{https://doi.org/10.1145/775152.775160}

\bibitem{DBLP:journals/ngc/KowalskiS86}
Kowalski, R.A., Sergot, M.J.: {A Logic-based Calculus of Events}. New
  Generation Computing  \textbf{4}(1),  67--95 (1986). \doi{10.1007/BF03037383}

\bibitem{kozlenkov06edbt}
Kozlenkov, A., Pe{\~n}aloza, R., Nigam, V., Royer, L., Dawelbait, G.,
  Schroeder, M.: Prova: Rule-based java scripting for distributed web
  applications: A case study in bioinformatics. In: EDBT Workshops. LNCS,
  vol.~4254, pp. 899--908. Springer (2006)

\bibitem{DBLP:journals/tplp/LukacsyS09}
Luk{\'{a}}csy, G., Szeredi, P.: Efficient description logic reasoning in
  prolog: The dlog system. Theory Pract. Log. Program.  \textbf{9}(3),
  343--414 (2009). \doi{10.1017/S1471068409003792},
  \url{https://doi.org/10.1017/S1471068409003792}

\bibitem{Manthey:Bry:SATCHMO:88}
Manthey, R., Bry, F.: {SATCHMO: a theorem prover implemented in Prolog}. In:
  Lusk, E., Overbeek, R. (eds.) Proceedings of the 9\/$^{th}$ Conference on
  Automated Deduction, Argonne, Illinois, May 1988. Lecture Notes in Computer
  Science, vol.~310, pp. 415--434. Springer (1988)

\bibitem{10.1007/978-3-540-30475-3_38}
Motik, B., Sattler, U., Studer, R.: Query answering for owl-dl with rules. In:
  McIlraith, S.A., Plexousakis, D., van Harmelen, F. (eds.) The Semantic Web --
  ISWC 2004. pp. 549--563. Springer Berlin Heidelberg, Berlin, Heidelberg
  (2004)

\bibitem{DBLP:journals/jair/MotikSH09}
Motik, B., Shearer, R., Horrocks, I.: Hypertableau reasoning for description
  logics. J. Artif. Intell. Res.  \textbf{36},  165--228 (2009).
  \doi{10.1613/jair.2811}, \url{https://doi.org/10.1613/jair.2811}

\bibitem{PRZYMUSINSKI1988193}
Przymusinski, T.C.: Chapter 5 - on the declarative semantics of deductive
  databases and logic programs. In: Minker, J. (ed.) Foundations of Deductive
  Databases and Logic Programming, pp. 193 -- 216. Morgan Kaufmann (1988).
  \doi{https://doi.org/10.1016/B978-0-934613-40-8.50009-9},
  \url{http://www.sciencedirect.com/science/article/pii/B9780934613408500099}

\bibitem{dlvhex-journal:2016}
Redl, C.: The dlvhex system for knowledge representation: Recent advances
  (system description). Theory and Practice of Logic Programming  \textbf{16}
  (07 2016). \doi{10.1017/S1471068416000211}

\bibitem{Sakama:PossibleModelSemanticsDisjunctiveDatabases:DOOD:89}
Sakama, C.: {Possible Model Semantics for Disjunctive Databases}. In: Kim, W.,
  Nicholas, J.M., Nishio, S. (eds.) Proceedings First International\
  Conference\ on Deductive and Object-Oriented Databases (DOOD-89). pp.
  337--351. Elsevier Science Publishers B.V.\ (North--Holland) Amsterdam (1990)

\bibitem{Sakama:Inoue:PossibleModels:JAR:94}
Sakama, C., Inoue, K.: {An Alternative Approach to the Semantics of Disjunctive
  Logic Programs and Deductive Databases}. Journal of Automated Reasoning
  \textbf{13},  145--172 (1994)

\bibitem{scala}
{The Scala Programming Language}, \url{https://www.scala-lang.org}

\bibitem{DBLP:conf/lpnmr/SyrjanenN01}
Syrj{\"{a}}nen, T., Niemel{\"{a}}, I.: The smodels system. In: Eiter, T.,
  Faber, W., Truszczynski, M. (eds.) Logic Programming and Nonmonotonic
  Reasoning, 6th International Conference, {LPNMR} 2001, Vienna, Austria,
  September 17-19, 2001, Proceedings. Lecture Notes in Computer Science,
  vol.~2173, pp. 434--438. Springer (2001). \doi{10.1007/3-540-45402-0\_38},
  \url{https://doi.org/10.1007/3-540-45402-0\_38}

\end{thebibliography}

\end{document}